# Selection and Coalescence in a Finite State Model

Irwin Kuntz[1]

January 4, 2017

**Abstract**

To introduce selection into a model of coalescence, I explore the use of modified integer partitions that allow the identification of a preferred lineage. I show that a partition-partition transition matrix, along with Monte Carlo discrete time kinetics, treats both the neutral case and a wide range of positive and negative selection pressures for small population sizes. Selection pressure causes multiple collisions per generation, short coalescence times, increased lengths of terminal branches, increased tree asymmetry, and dependence of coalescence times on the logarithm of population size. These features are consistent with higher order coalescences that permit multiple collisions per generation. While the treatment is exact in terms of the simplified Wright-Fisher model used, it is not easily extended to large population size.

[1] Department of Pharmaceutical Chemistry, University of California, San Francisco, Box 2280, 94158-2517
E-mail: kuntz@cgl.ucsf.edu

# 1 Background and Motivation

Neutral coalescent theory has offered elegant insights into the structure and dynamics of populations on a time scale of generations (Kingman, 1982; Hein et al., 2005; Wakeley, 2009), on a time scale of generations. Direct incorporation of selection has been pursued since Haldane, 1927 and Fisher, 1937, but adding selection directly into the Kingman coalescent has proven difficult, presumably because of the assumptions of exchangeability of parents and the limitation to a single coalescence per generation in the Kingman model are inconsistent with most models of selection.

Selection has been explored in a variety of systems (see review by Wakeley, 2010). There have been a number of studies of 2 allele systems (e.g. Wilkinson-Herbots, 2008; Uecker & Hermisson, 2011). An ancestral graph method retaining the Kingman assumptions has been proposed (Krone & Neuhauser, 1997). There have been a number of papers that address coalescence of adapting populations using "traveling wave" equations (Tsimring et al. 1996; Brunet & Derrida, 1997; Rouzine et al., 2003; Brunet et al., 2007; Desai et al., 2013).

A major step towards a more general theory was the realization that the Kingman coalescent is a member of a larger class of coalescent processes. Some members of the family (e.g. the $\Lambda$ and $\Xi$ coalescents) allow for multiple coalescent events per generation while retaining the assumption of equivalence of individuals (Pitman, 1999, Sagitov, 1999). Simultaneous multiple collisions are allowed in the $\Xi$ coalescent (Schweinsberg, 2000). A specific model from applied physics has been of particular interest (Bolthausen & Snitzman, 1998) and there are now many studies of applying such coalescent processes to population genetics (Möhle & Sagitov, 2001; Brunet et al., 2007; Neher & Hallatschek, 2013; Tellier & Lemaire, 2014; Bah & Pardoux, 2016). These treatments are typically focused on the limiting case of large population size and low selection pressures.

In this paper I explore a finite state model with small populations where it is possible to enumerate all outcomes. Selection can be included directly with explicit consequences for coalescence. This formulation is useful over a wide range of selection pressure and demonstrates a continuous shift from Kingman-like to multiple collision phenomena as the selection pressure increases. Exchangeability of parents is removed by designating a specific allele or lineage associated with a selective advantage. For numerical reasons, the finite state approach in this form is limited to small population sizes but it provides useful insights into general coalescent processes.

I use two types of calculations, both based on a modified Wright-Fisher model (Wright, 1931; Fisher, 1930; Nielsen & Slatkin, 2013). I follow the coalescence of a collection of diverse haploid alleles or lineages in a population of constant size. By lineage I simply mean a set of individuals related by common ancestry. An individual entity can persist and reproduce for more than one generation, a so-called "over-lapping generation" model. Individuals have binomial-distributed numbers of offspring per generation with the births balanced by deaths. Other birth-death protocols are readily studied. No mutation or migration takes place. Progeny inherit parental genomes without error and maintain those genomes and fitness unchanged throughout their lives. Lineages cannot interconvert and there is no recombination. Mutation and migration processes can be added later, but they do not play into the main results of this paper.

The first calculation employs partitioning methods (see Pitman, 1999 and Methods Section) to keep track of the distribution of lineages. I calculate a transition matrix for conversion of one distribution into another. To study coalescence, I focus on a discrete time Markov process in which multiple initial lineages are reduced to a single lineage. In the absence of selection pressures and without mutation or migration processes, the population undergoes neutral drift until all members of the population belong to the same lineage.

A second approach directly simulates the discrete time evolution of the Wright-Fisher model from a set of N diverse individuals to coalescence. This process is repeated a large number of times to gather statistical data. The two approaches agree quantitatively within the statistical errors of the direct simulation.

In either approach, I calculate the mean coalescence time, $\tau$, its distribution, and its variance as a function of the number of initial entities, **N**, and as a function of a relative selection variable, **s**, which endows a particular lineage with an improved chance of reproduction. The calculations of the lineage population distributions are exact in terms of the finite state model assumptions. It is also possible to infer many properties of the coalescent trees such as collision frequencies, relative branch lengths, node and tree asymmetries and their dependence on selection pressure.

In this study, selection alters the odds of reproduction of the selected lineage. These changes are propagated through the transition matrix or introduced into the reproduction phase of the Wright-Fisher simulation. Other ways of introducing selection could be examined (e.g. Bah & Pardoux, 2016).

It is important to note that, to study selection, I modify the usual partition representation by ordering the component parts by lineage. Without this change, the partition methodology can be applied to neutral processes but it will not provide a useful route to selection.

While the finite state model as used here should be considered a "toy" model because of its simple approach to selection and the limits on population size, it offers a straightforward way to study coalescence under selection.

## 2 Methodology

### 2.1 Mapping of Distributions of Lineages onto Partitions of Integers.

I map distributions of lineages onto partitions of integers. Integer partitioning comes directly from number theory and has a rich history in population genetics being embedded in the Ewens Sampling Formula (Ewens, 1972; Kingman, 1978; Wakeley, 2009). Pitman makes use of integer partitions in his study of coalescents with multiple collisions (Pitman, 1999). There have been many papers on such processes (e.g. Goldschmidt & Martin, 2005; Drmota et al, 2007; Freund & Siri-Jégousse, 2014).

An accessible treatment of partitions is available (Andrews & Eriksson, 2004). In brief, a partition of a positive integer, N, is a set of positive integers, called "parts", whose sum is N. For values of N > 1, there are multiple partitions. For example, for N = 4, there are five partitions. The number of partitions of a given integer is called the "partition number", $\mathcal{P}$ **(N)**. It can be calculated in asymptotic or recursive form, but the formula is not needed for this work. Efficient computer algorithms exist for generating all partitions for a given integer (Kelleher & O'Sullivan, 2014). Generally, the position of a part has no significance, and "parts" of the partition are arranged in either ascending or descending order with respect to the largest integer in the partition as a matter of convenience. For our purposes, these two representations are equivalent and either serves to study neutral coalescence. However, to study selection, I need to order the parts in a different manner.

Specifically, I formally assign a label to each kind of lineage so that the parts count the number of individuals in a current population that share a common ancestor. As in the Ewens sampling formula, if a particular kind of label is not represented in the partition, a zero is entered. However, in the partitions used here, labels are ordered so that a particular lineage is found at a particular position in the partition. These partitions can be thought of as an **N**-element vectors whose elements sum to **N** with zeroes interspersed as needed. In the combinatorics literature, a partition with a particular ordering of the parts has been called a "composition" and with added zeros, it is called a "weak composition" (Heuback & Mansour, 2004). It then becomes possible to get a complete description both for the neutral case and when selection is present. I will call these partly ordered partitions "sub-partitions". There are $\mathcal{SP}$ **(N)** sub-partitions for the integer **N** (see Table 1).

| Integer $\mathcal{N}$ | Partition Number $\mathcal{P(N)}$ | Number of Sub-partitions $\mathcal{SP(N)}$ | Explicit permutations $\mathcal{EP(N)}$ |
|---|---|---|---|
| 1 | 1 | 1 | 1 |
| 2 | 2 | 3 | 3 |
| 3 | 3 | 6 | 10 |
| 4 | 5 | 11 | 35 |
| 5 | 7 | 18 | 126 |
| 6 | 11 | 29 | 462 |
| 7 | 15 | 44 | 1716 |
| 8 | 22 | 66 | 6435 |
| 9 | 30 | 96 | 24310 |
| 10 | 42 | 138 | 92378 |

Table 1: Partition, Sub-partition Numbers and numbers of explicit permutations. Partition Numbers from the literature (Abramowitz & Stegun, 1964) and directly from computation (Kelleher & O'Sullivan, 2014). The sub-partition numbers were obtained from the computations in this paper. There is a simple recursive function for sub-partition numbers: $\mathcal{SP}(\mathbf{N}) = \mathcal{SP}(\mathbf{N-1}) + \mathcal{P}(\mathbf{N}).$ The numbers of explicit permutations, $\mathcal{EP(N)},$ can be calculated from the permutations of the parts of each partition with the partitions expanded into "weak compositions" through the addition of sufficient zeros to provide N elements for each composition. Then, $\mathcal{EP}(\mathbf{N}) = \sum \mathbf{N!}/\prod \mathbf{M_i!}$ with the sum over all the partitions of N and the $M_i$'s are the numbers of each integer appearing in the composition, including the zeros

My reason for mapping lineages to partitions is to count the number of lineage arrangements to obtain the probability of transition of one partition into another under a rule set that defines operations such as selection, reproduction and population control. Without selection, a set of lineages will coalesce into a single lineage under a process called "neutral drift", with the "winning" lineage being a matter of chance. The time scale of this process is accurately described by the Kingman coalescent for moderate to large population sizes for the Wright-Fisher model {Kingman, 1982} and the corrections at small population sizes are well understood {Wakeley, 2009}. Adding selection into the same framework is one of the main motivations for using the finite state model.

**2.2 Transition Matrix Elements.**

I need to calculate the elements of a transition matrix for the conversion of one partition or sub-partition into another. This process can be thought of as counting the number of hands of cards obeying certain rules that can be drawn from a deck of known composition. I describe the method of calculating the transition elements for a system under selection. The neutral system elements can be obtained by setting the selection coefficient, **s**, to zero.

To calculate the elements in a consistent manner, I designate one sub-partition of a pair as the "originating" sub-partition and the matrix element is the probability of its conversion to a "target" sub-partition.

**2.3 Selection.**

Incorporation of natural selection into coalescent theory has been a major goal of population genetics theory. Wakeley has an excellent review of the complex issues involved (Wakeley, 2010). A strong point of the finite state model is that selection can be introduced in a simple way across a wide range of selection pressures. A selected lineage is given an exponential change in fitness of 1+**s** over the rest of the lineages, where **s** is the selection coefficient. The partition transition probabilities are then calculated. In effect, I use a "weighted" deck of cards, with the chosen lineage getting an advantage by having the equivalent of more cards in the deck. The results are normalized across each row of the transition matrix with only the relative transition probabilities being required. In the model used here, selection increases the chances of reproductive success. Bah & Pardoux (2016) describe a selection based on enhanced death processes for non-selected lineages.

We can explore a wide range of values of $s_a$, including negative values (i.e. lineages that are at a selective disadvantage). As noted, this approach is a "toy model" of selection. It is much simpler than the many models of selection defined for sequences of multiple genomic sites (e.g., Wakeley, 2010; Neher & Hallatschek, 2013). However, as I show, the model reproduces the main features of more elaborate treatments. Formally, this same approach can be extended to systems where more lineages are undergoing selection or there is a distribution of selection pressure is available to individuals within a lineage. These enhancements would require a more complex organization at the sub-partition level. Another way to explore such systems is to use a simulation technique described later.

Assume one lineage is at selective advantage with respect to all other lineages which are selectively neutral to each other: $s_a$ = **s**, $s_b$ = $s_c$ ....= **0.** Transition matrix elements depend on **s** in polynomial fashion. The absolute transition elements increase with increasing values of **s**. In contrast, the normalized transition elements can increase or decrease as **s** increases.

As noted, the counting procedure can be readily visualized as a card game. Three steps are needed. The first step is to count target "hands" that are permutations of the order of the draw of the cards. The second step is to deal with the change in odds due to selection pressure. The third step is a formula for evaluating the odds of all the hands that do not contain the selected lineage label.

I calculate the transition matrix elements, $T_{ij}$ on an incremental hand-by-hand basis using the product of the three steps in **Eqn. 1.** Each hand is assigned to a sub-partition using a simple hash code and the relevant transition matrix element, $T_{ij}$.

$$\Delta T_{ij} = Per_{ij} * S_{ij} * O_{ij} \qquad Eqn.\ 1$$

The first term on the right accounts for permutations of the labels themselves. There is nothing unique about a lineage label, so labels for any of the **N** lineages can be found in the hand and there can be more than one copy of a given lineage. The formula for the number of permutations with repetitions is given in **Eqn. 2,**

$$Per_{target\ j} = \frac{N!}{\prod_{k=1}^{k=N} P_{j,k}!} \qquad Eqn.\ 2$$

where the product is over the number of copies of each label in the hand for the target partition. The numbers of copies of each label are just the parts of the partition, $P_j$. For example, for a system of 6 lineages: there are **N!** = 720 ways that the labels a,b,c,d,e,f can be chosen; while for target hands containing a,a,b,c,d,e in any order there are **N!/2!** = 360 permutations.

The next step calculates the contribution of a hand with $P_{i,1}$ copies of the selected lineage in the target sub-partition, $P_{j,1}$ copies of the selected lineage in the originating sub-partition and the selection coefficient **s** (**Eqn. 3**).

$$S_{i,j} = [(1+s) * P_{i,1}]^{P_{j,1}} \qquad Eqn.\ 3$$

The number of target hands that arise from the relabeling of the parts given the deck of cards for each of the originating partitions is given in **Eqn. 4**. The probability of each representative target hand is computed conditioned on the composition of the originating sub-partition.

$$O_{i,j} = \prod_{k=2}^{k=N} (P_{i,k})^{P_{j,k}} \qquad Eqn.\ 4$$

Here the product is taken over all parts, **k = 2,N**. Again, $P_{i,k}$ refers to the parts of the originating sub-partition, and $P_{j,k}$ refers to the parts of the target sub-partition.

The appropriate element in the transition matrix is incremented as shown in **Eqn. 1** and when the process is complete, a normalized transition matrix, $T^{\#}$ is calculated from $T_{ij}$ with a row-by-row normalization to yield row sum probabilities of 1. The transition matrices used in this work were based on sub-partition to sub-partition conversions. To obtain the neutral case, the selection coefficient is set to **s = 0** in **Eqn. 3**.

A related calculation is found in Wakeley (Wakeley, 2009, Section 3.2.1) where, for the neutral system, he obtains the probabilities for certain <u>collections</u> of partitions transforming into other <u>collections</u>. Specifically, he groups partitions that contain the same number of lineages together. For example, {3,1} and {2,2} are both partitions of 4 that contain 2 lineages. In Wakeley's calculation, these two partitions are combined. He presents an efficient way of carrying out the permutations for these combinations using Stirling numbers (Abramowitz & Stegun, 1964).

Keeping the full transition matrix carries benefits in sorting out the time course of coalescence, and more importantly, it provides a route to selectivity, not available from Wakeley's treatment. Thus, by setting up a lineage-sorted ordering to partitions and explicit enumeration, we can generate both a complete list of all distributions of lineages **and** a natural way to introduce selection.

**2.4 Markov Model for Coalescence.**

Coalescence is treated as a stochastic process in forward time representing a decrease in the number of ancestors of a current population of lineages (**Fig. 1**). Coalescence to the most recent common ancestor occurs when all partitions except {N} have zero occupancy. We will see that the coalescent process for the selected lineage has many of the features of the Bolthausen-Sznitman coalescent (Bolthausen & Sznitman, 1998; Pitman, 1999; Brunet et al., 2007; Neher & Hallatschek, 2013; Tellier & Lemaire, 2014).

The coalescence calculations were carried out using the discrete time Markov process (**Eqn. 5**). I assume that the transition matrix elements are time invariant and use **Eqn. 5** to obtain the occupancies of all partitions, $P_i$, or sub-partitions, $SP_i$, for each generational step (see Pitman, 1999). The initial state has the entire population in the most diverse partition. The final partition in the ascending formulation, {N}, represents the coalesced population. In the absence of mutation and migration, this partition represents an "absorbing" state and its relative occupancy accumulates until all other lineages have been removed. The time dependence of the occupancy of the terminal partition is the summation of all the paths leading to coalescence. The numerical derivative of this time dependence provides the distribution of coalescence times (Tavaré, 1984;Takahata & Nei, 1985) from which the mean, variance and other properties are readily obtained.

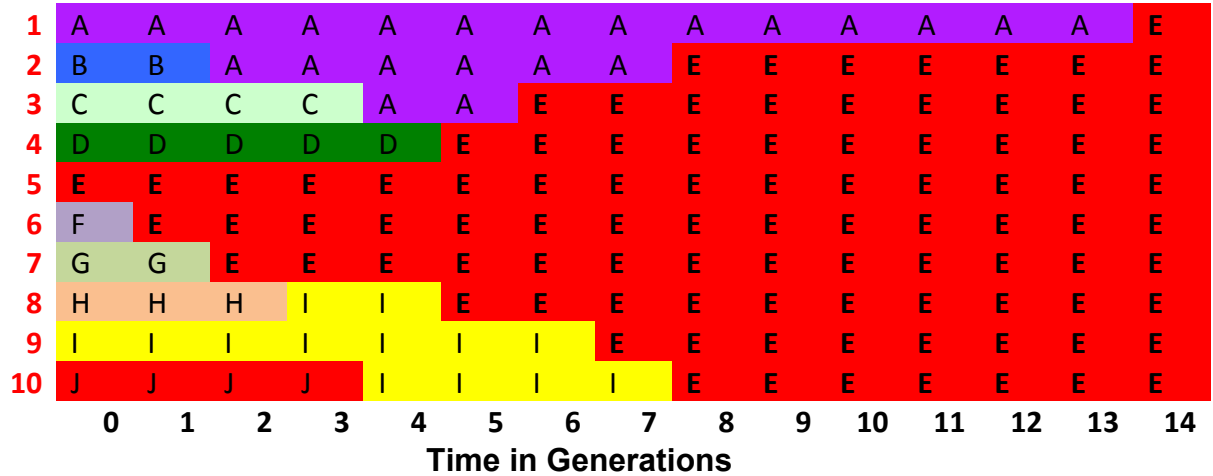

Figure 1. Coalescent Process: Time runs from the left in generational steps from Generation 0 to coalescence at Generation 14. At t = 0, there are 10 individuals each of a different lineage. The lineages are labeled A-J. Without selection, the labels are interchangeable. In each generation, the population size is limited to N = 10. Some lineages expand and others contract or go extinct on a random basis. No new lineages can be introduced since there are no mutation or migration mechanisms. Ultimately, only one lineage survives at t = 14, and the system has coalesced. Lineage E is the most recent common ancestor for the individuals in generation 14. Because I track lineages rather than individuals, this diagram represents a large number of coalescent trees. Partitions at each time point can be obtained by summing the number of individuals in each lineage at each generation (see text). A very similar figure for alleles is found in Desai et al., 2013.

$$X_i(t) = \sum_{j=1}^{j=P(N)} X_j(t-1) * T_{ij}^{\#} \qquad Eqn. 5$$

One caution: the discrete time representation used in this work limits the time resolution and hence the accuracy of the coalescence times when their absolute values drop below ca. 3 generations.

These calculations are readily programmed and the probabilities for neutral systems match exactly those generated from exhaustive enumeration. To confirm that the formulas are correct when selection is present, I also set up simulations that determined the transition elements statistically. The simulations check both the odds calculations and the Markov kinetic procedure.

**2.5 Simulations.**

I use a discrete time simulation that incorporates the same focus on lineage propagation and selection effects. At each time step, the resulting lineage distribution is mapped to a sub-partition and is used to increment the specific sub-partition occupancy, sub-partition to sub-partition transition matrix element, and, should coalescence be achieved, the coalescent time distribution. The sub-partition statistics are accumulated over a large number of iterations and compared to the formulaic solutions (**Eqns. 1-5**). The results, below, agree with the permutation computations within statistical error. The simulations also suggest new applications for the finite state model.

The simulation uses **M** multiple repetitions of a fixed population of **N** individuals. Each instance is followed, in turn, until its population has coalesced to a single lineage. The coalescence times and the sub-partition, sub-partition transitions are tabulated as they occur. After the full set of **M** coalescent events, an ensemble average is taken. Again, a useful heuristic for the simulations is a game of cards. The important aspects of the game are: the "deck" from which cards are taken, the original hand that is drawn, the expansion of the hand to mimic "reproduction", and the population control module that operates on the expanded hand to restore a new hand corresponding to the desired population size. This procedure is repeated until coalescence occurs. The deck consists of a number of "suits", one suit per lineage. The suits are labeled, say, "a", "b", etc. with no significance ascribed to a particular label. With no selection, the deck would consist of equal numbers of cards of each label. Selection is readily introduced as a change in the number of cards for a selected lineage. Cards are drawn with replacement. For each repetition, the initial hand consists of one card of each lineage. Other initial conditions can be explored as desired. The "reproduction" phase requires drawing some number of cards that match each card in the hand. The current model adds 1 new card for each card of that label. Other reproduction paradigms are easily created with smaller or larger numbers of new cards distributed in some fashion.

To summarize: at the end of the reproduction phase, a new hand has been generated consisting of N "parent" cards and N progeny cards or 2N total cards. None of these cards was chosen by chance, each was determined by the game protocols.

For the population control step, a card in the reproduction hand is selected at random. A card from the deck with the same label is placed in the new hand. This procedure is repeated until the new hand contains the requisite number of cards. It does not matter if a card from the reproduction hand is selected more than once since the label, rather than the card itself, is transferred to the new hand each time it is selected.

Once the new hand is completed, its sub-partition is calculated and the element in the transition matrix indexed by the {original sub-partition, target sub-partition} pair is incremented. A record is also kept of the time the target sub-partition was generated. As the process repeats, a time point is reached where a hand contains only a single suit. The population has "coalesced" and this time point is entered into the distribution of coalescence times.

The procedure is repeated **M** times. As expected, the variance in the coalescence times fall as the sampling increases (**Fig. 2**). As can be seen, the simulation results converge to the formulaic calculations within statistical error.

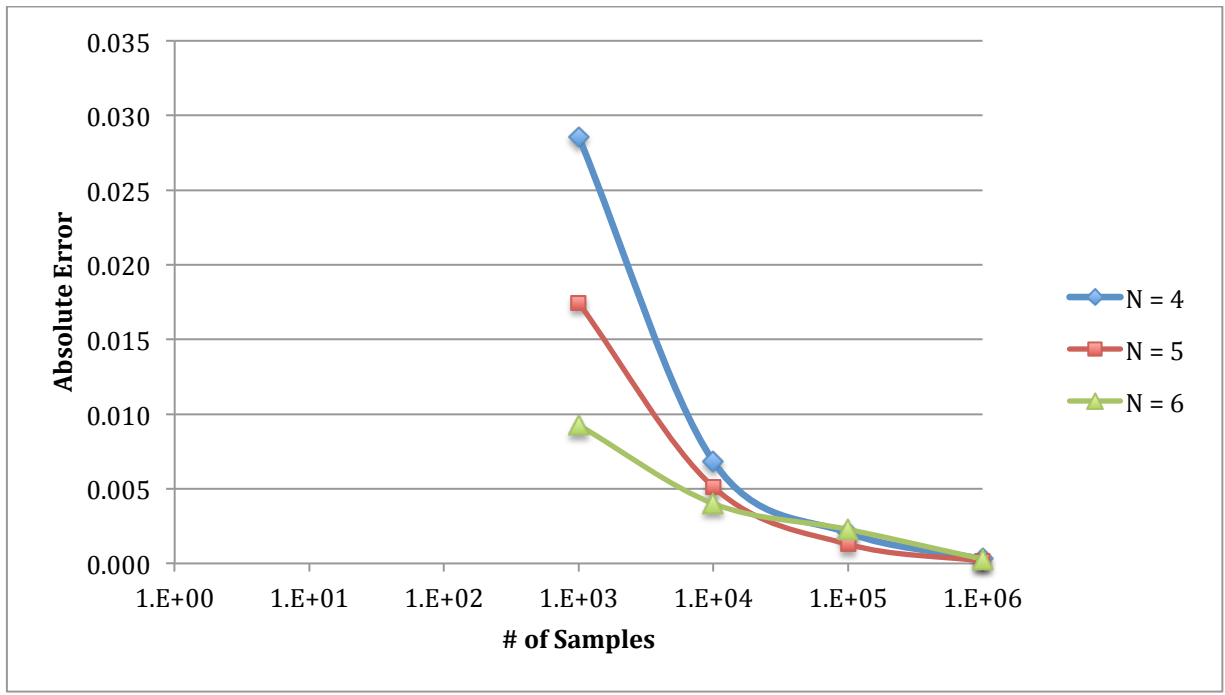

Figure 2. Convergence of Coalescence Times obtained from simulation procedure is compared to times calculated from permutations as a function of number of samples and population size.

Selection can be added to the card game in many ways. The make-up of the deck can be altered. The reproduction protocol could reflect a selection criterion. The population control (survival) protocol could be modified. I choose to alter the card deck itself by adding or subtracting cards to the lineage under positive (or negative) selection pressure. For positive selection, an allocation of **1+s** cards was added to the reproduction hand. If **s** were not an integer, the total number of cards was scaled up so that an integer number of cards could be added. For negative selection, cards were subtracted from the reproduction hand. Again, a scale factor was introduced to guarantee integer manipulations without changing the odds.

The agreement between observed and calculated coalescence times under selection was found to be within the sampling error, suggesting that the rule set outlined above produced the same odds as the formulas did.

Beyond the obvious desire to confirm the calculations, there is an advantage to the simulation version. It immediately offers ways to expand the finite state approach to other topics of evolutionary interest. There are direct ways to provide distributed reproductive or selection strategies for more than one lineage. One could explore reproduction and survival with different selection/fitness criteria. Adding mutation or migration can be readily accomplished as well. For example, mutation can be added by providing an operation that converts a given card to a new lineage. The new lineage could be transferred into the card game based on rules about its selective advantage, its reproduction rate, etc. and its fate could be followed either as a random or deterministic event. Migration among islands could be modeled in a similar fashion.

In sum, these games provide excellent metaphors for many evolutionary processes.

The programs were written in the GNU version of Fortran 90. They take only a few minutes of clock time on a HP DL580 G7 server using a single processor. Source code is available from the author.

**2.6 Summary of Calculations.**

I present a "finite states" representation of the distribution of lineages coupled with a Markov calculation that provides lineage distributions as a function of population size, generation time, and selection pressure. The main advantage of the approach is the transparency of the effects of selection. A second advantage is obtaining in a direct way the full distribution function for the complete set of coalescence paths or any desired subset of the paths. The main disadvantage is that the rapid growth of the internal states. I emphasize that the numerical results in the next sections are derived from a particular birth-death model. However, the methodology can be extended to other models, including models having more than one selected lineage. As shown in the Results, in most respects, the finite state model under selection pressure generates the features expected for enhanced coalescent processes that permit multiple mergers per generation (Neher & Hallatschek, 2013; Tellier & Lemaire, 2014).

# 3 Results

The results are organized in the following order: 1) the finite state neutral coalescent; 2) effects of selection on coalescent times including the effects of negative selection; 3) selection effects on the properties of coalescent trees. The conclusion is that the neutral finite state produces a Kingman-like coalescent. Selection causes marked deviations in many properties that closely resemble the behaviors displayed in systems experiencing a multiple merger coalescent.

**3.1 The Neutral Coalescent.**

To obtain the distribution of coalescence times, I use a normalized matrix of partition-partition transition probabilities, calculated as described in the Methods section. Coalescent distribution times for $N = 6$ are compared to that obtained for a Kingman coalescent (Tavaré, 1984) in **Fig. 3.** The two approaches produce similar but not identical results. The Tavaré-Kingman has a somewhat longer tail. The means, modes, and standard deviations are within a few percent of each other (**Table 2 & Fig. 3B**). The standard deviations reach limits as **N** increases: 1.16 for the Kingman model, 1.02 for the finite state model.

These small differences are expected because Kingman's coalescent is calculated in continuous time (Nielsen & Slatkin, 2013) and allows only transitions where the number of lineages either does not change or decreases only by 1. In addition, Kingman's derivation applies in the limit of an infinite population size. In contrast, the Markov process for the Wright-Fisher model is in discrete time (Nielsen & Slatkin, 2013) and I deliberately examine small values of **N**.

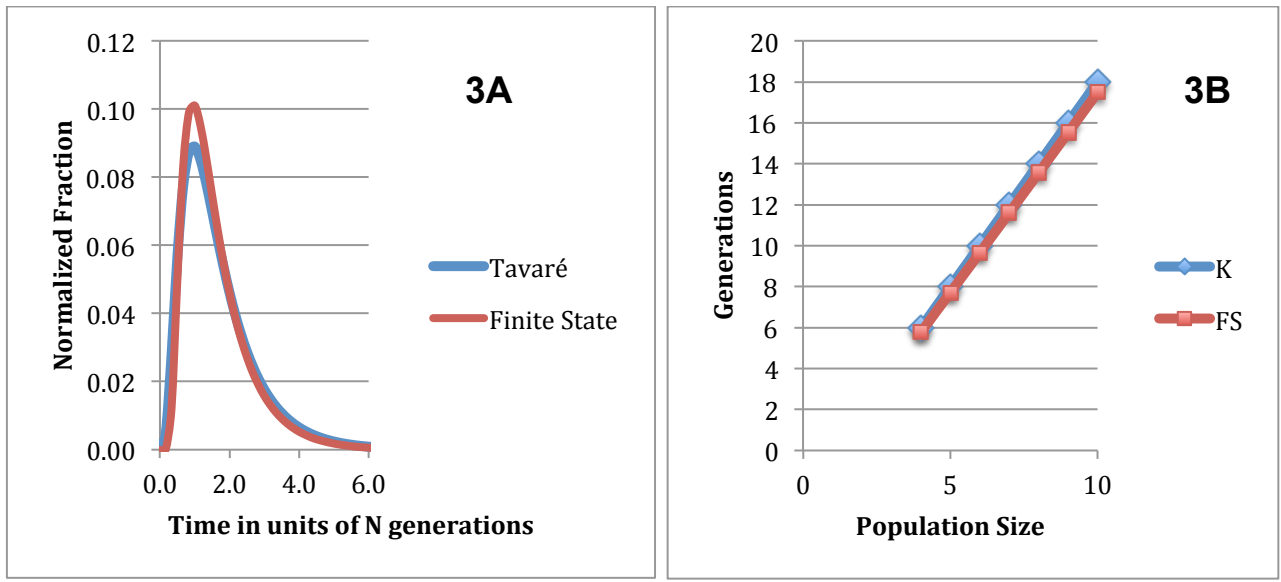

Figure. 3. The Finite State Neutral Coalescence Time Distributions and Dependence on Population Size are very similar to the Kingman Coalescent calculated from Tavaré (1984). Fig. 3A Distribution for N = 6 Finite State method (red), Kingman (Blue) for N = 6. The curves are normalized so that the probabilities sum to 1. Fig. 3B: Near quantitative agreement for the mean neutral coalescent times calculated from Kingman and Finite State Models as a Function of Population Size.

| N | $\tau_{FS}$ | $\tau_K$ | Std. Dev.$_{FS}$ | Std. Dev.$_K$ |
|---|---|---|---|---|
| 4 | 1.45 | 1.50 | .91 | 1.14 |
| 5 | 1.54 | 1.60 | .95 | 1.15 |
| 6 | 1.61 | 1.67 | .97 | 1.16 |
| 7 | 1.66 | 1.71 | .99 | 1.16 |
| 8 | 1.70 | 1.75 | 1.00 | 1.16 |
| 9 | 1.73 | 1.78 | 1.01 | 1.16 |
| 10 | 1.75 | 1.80 | 1.01 | 1.16 |

Table 2. Means and Standard Deviations from the Finite State (FS) and Kingman (K) models in Units of N generations.

## 3.2 Selection.

For selection, I use of a "weighted" deck of cards with more cards of the selected suit (lineage) than the other suits. The probability of drawing the suit associated with the preferred lineage (**a**) is increased by $(1+s_a)$ where $s_a$ is the selection advantage for that lineage. We can explore a wide range of values of $s_a$, including negative values (i.e. lineages that are at a selective disadvantage).

Assume one lineage is at selective advantage with respect to all other lineages which are selectively neutral to each other: $s_a = s$, $s_b = s_c \ldots = 0$. Transition matrix elements depend on **s** in polynomial fashion. The absolute transition elements increase with increasing values of **s**. In contrast, the <u>normalized</u> transition elements can increase or decrease as **s** increases. **The net effect is that selection increases the probability of multiple coalescences per generation and reduces the number of lineages at a given generation at constant population size**, leading to a reduction in the mean coalescence time (**Fig. 4**).

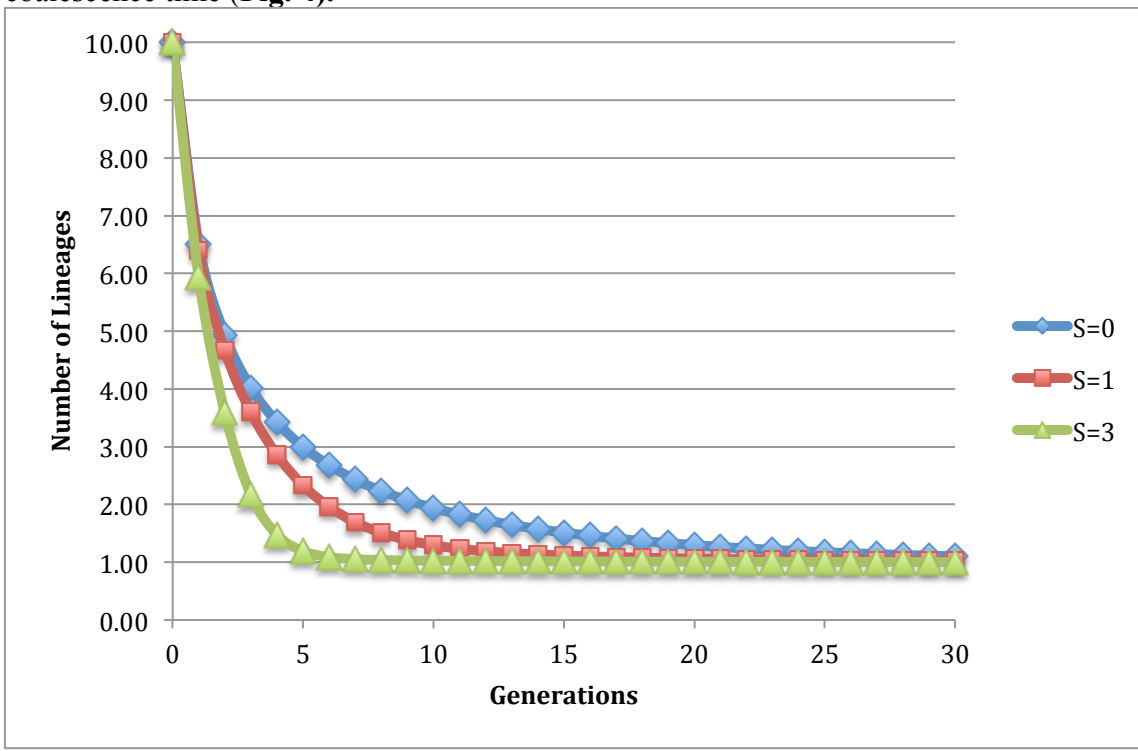

Figure 4. Positive selection pressure decreases the number of lineages as a function of time for N = 10. The Kingman coalescence process is indistinguishable from s = 0 at this resolution.

For a system where a single lineage has a fitness advantage, we can examine times to three different end states: the time it takes the favored lineage to sweep to coalescence, the time it takes an un-favored lineage to coalesce to an un-favored final state, and the time it takes for the total population to reach coalescence to a common ancestor (Fig. 5A). For the selected lineage, mean coalescence times fall slowly at low **s** and then more rapidly as **s** increases, approaching a limit of **1/s** at the highest **s** values (**Fig. 5B**). Interestingly, coalescence times of the unselected lineages **(Fig. 5A)** show almost no effect of the selection pressure, presumably because, except for the initial state, enough partitions containing only unselected lineages exist to lead to coalescence. The selection pressure dependence of the coalescence for the selected lineage and the total population is complex. Empirically, the full finite state data are better described by a functional form $\tau(s)/t_0 = 1/(1+s)$ (orange curve, **Fig. 5B**). However, at high **Ns** regimes, we find $\tau$ approaching $1/s$ (violet curve, **Fig. 5B**) in agreement with others (Wakeley, 2010; Neher & Hallatschek, 2013; Desai et al, 2013). The form of the transition elements from the finite state model is a ratio of polynomials, and no simple functional form quantitatively covers the entire selection range.

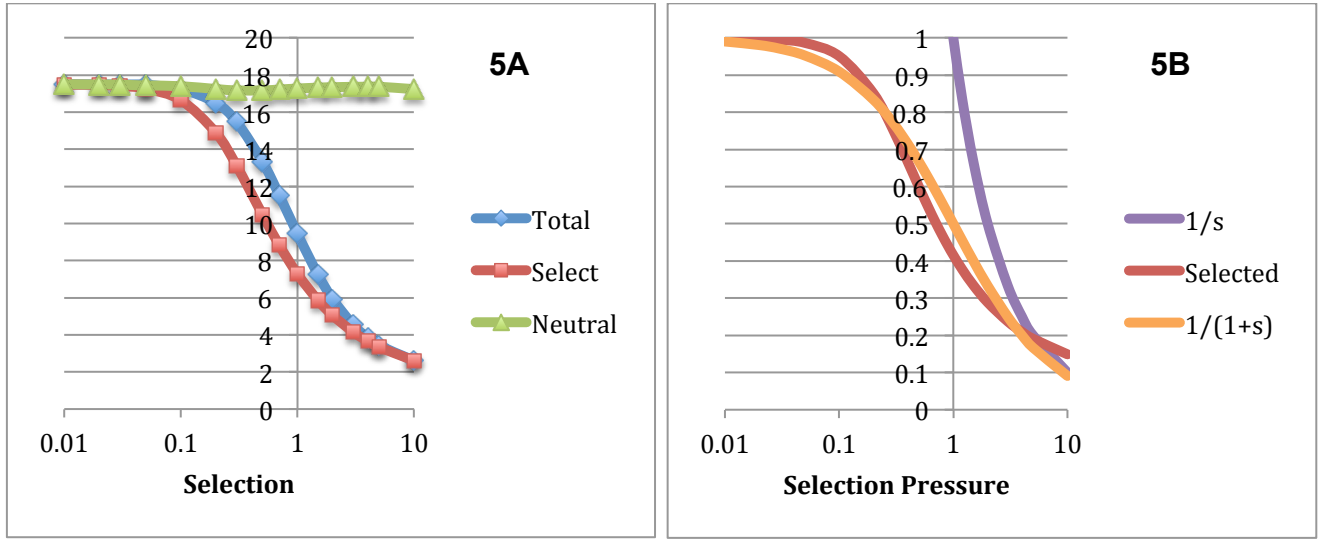

Figure 5. Panel 5A. Selection pressure causes a large reduction in Mean Coalescence Times measured in generations (Y-axis) for the selected lineage (blue), with almost no effect on unselected lineages coalescing to an unselected end state (green). The mean coalescence time for the total population (red) is just a population-weighted average of the green and blue curves. N = 10. Panel 5B. The observed dependence of Mean Coalescence Time for the selected lineage (red curve) is better correlated with 1/(1+s) (green curve) rather than 1/s shown by the blue curve (Neher & Hallatschek, 2013.

Selection reduces the mean, mode and standard deviations (**Table 3)**. The distributions become more symmetrical as selection pressure increases, largely because the long tails are truncated. The coefficient of variation falls towards a limiting value of 0.25 at high selection pressure.

| Selection Pressure | Mean Coalescent Time | Std. Dev. | Coefficient of Variation |
|---|---|---|---|
| 0.00 | 17.47 | 10.11 | 0.58 |
| 0.05 | 17.26 | 9.93 | 0.58 |
| 0.20 | 14.88 | 7.93 | 0.53 |
| 1.00 | 7.23 | 2.66 | 0.37 |
| 2.00 | 5.05 | 1.58 | 0.31 |
| 3.00 | 4.15 | 1.21 | 0.29 |
| 5.00 | 3.34 | 0.91 | 0.27 |
| 10.00 | 2.61 | 0.67 | 0.26 |

Table 3. Coalescent properties as a Function of Selection Pressure in Generations. N = 10.

**3.3 Negative Selection of a Single Lineage.** Selection is often treated as a positive enhancement. However, the effects of mild negative selection form part of the neutral theory of evolution (Kimura, 1983). With the finite state model, it is possible to treat "negative" selection directly as a disadvantage to a specific lineage. The formalism sets the lower limit of **s** to -1.00 to avoid negative elements in the transition matrix. A plot of the mean coalescence times as a function of **s** from **s** = -.95 to 5.0 is given in **Fig. 6A** for N = 10.

The mean coalescent time is a maximum for **s** = 0. For both positive and negative selection, $\tau$ decreases. While the negative and positive selection arms of **Fig. 6A** look asymmetrical, this effect arises from using the selection coefficient as the x-axis. If I use the reciprocal fitness [$1/(1+s)$] for the negative arm, and the fitness $(1+s)$ for the positive arm, the two curves coincide **Fig/ 6B)**.

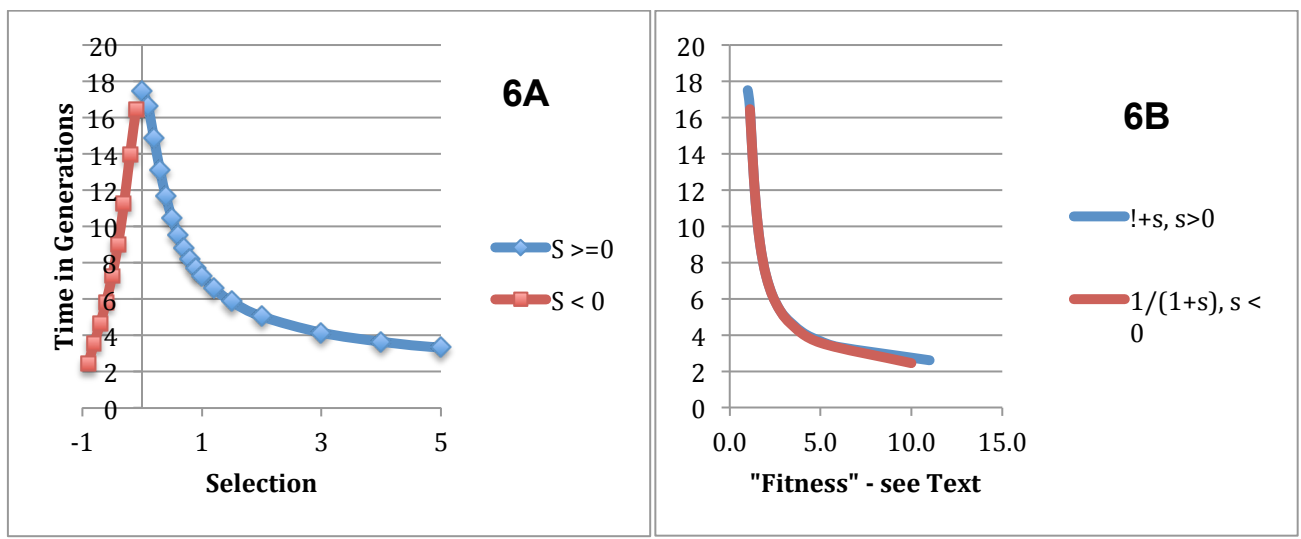

Figure 6. Negative Selection Pressure Reduces Mean Coalescence Times (red curves) for the selected lineage at N = 10. By changing the X-axis the positive selection (blue) and negative selection (red) curves are shown to be equivalent (see text).

Thus, for the selected lineage, coalescence under negative selection is very similar to coalescence under positive selection. If comparisons are made between pairs of selection coefficients that yield (roughly) the same coalescence times (e.g. **s** = -.5 and **s** = 1), the same partitions are involved in the same time sequences. The big difference is that the <u>occupancies</u> of the partitions of the selected lineage are vastly different. Positive selection acts to limit longer time contributions to the coalescent by expanding the number of multiple coalescences per generation while negative selection limits the longer time contributions by rapidly removing the (de)selected lineage from the population.

**3.4 Coalescent Times and Population Size.** At neutrality, the mean coalescent times are a linear function of population size (**Fig. 3B**). Nonlinear effects are observed as selection increases (**Fig. 7**). At high selection pressure and large population size, the coalescent time grows only as the logarithm of the population size or even as log log **N** (Neher & Hallatschek, 2013, Brunet & Derrida, 2013). The finite state results approach log **N** behavior over the population size range studied here. Such slow growth is characteristic of Bolthausen-Sznitman coalescents (Beretycki, 2009).

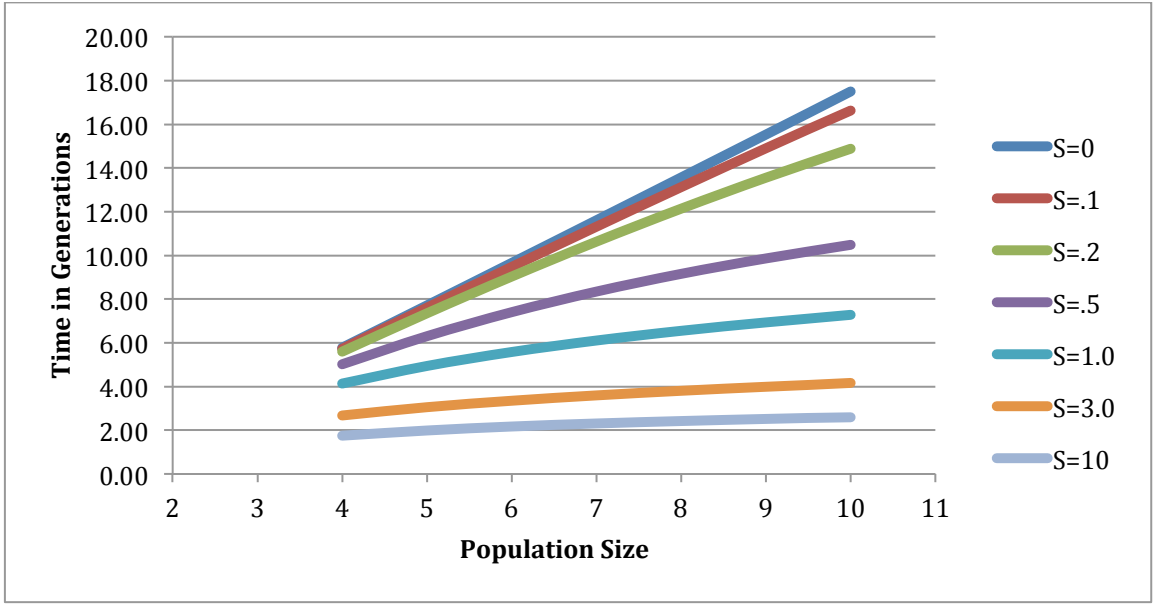

Figure 7. Coalescence times are a nearly linear function of population size at very low selection pressure (upper curves) and approach a logarithmic function of population size at high selection pressure (lower curves).

**3.5 Properties of Coalescent Trees Under Selection.** The finite state model with explicit selection-based sub-partitions allows the exploration of the average properties of trees as a function of time. The simulation calculations allow inspections of individual trees. Both approaches are useful. However, complete phylogenetic trees, with full descriptions of each individual's life span and progeny, remain a worthy goal for future studies (e.g. Figure 2 in Aldous, 1996).

Neher and Hallatschek apply the traveling wave model to study genealogies in large rapidly adapting haploid populations (Neher & Hallatschek, 2013). They discuss five features of coalescents formed in such systems that differ from the neutral Kingman predictions but are consistent with a Bolthausen-Sznitman or higher order coalescence process: 1) non-monotonic distribution of coalescences per generation, 2) smaller root to branch length ratios, 3) asymmetric or unbalanced trees with an uneven number of descendants at branch points, 4) non-monotonic site-frequency spectra, and 5) the non-monotonic distribution of pair-wise coalescence times. Related results are found in Desai et al., 2013, who also note clonal competition behavior before the coalescence is completed (Fig 1, Desai et al., 2013). We find many of these features in the coalescence behavior of the finite state model and can add some details to the Neher-Hallatschek observations.

**3.5.1 Multiple coalescences per generation.** A neutral system has a preponderance of zero or one coalescence per generation Wakeley, 2009). The direct effect of selection is to increase the chance of multiple coalescences per generation (Brunet et al. 2007, Neher & Hallatschek, 2013). I summarize these effects in **Fig. 8**.

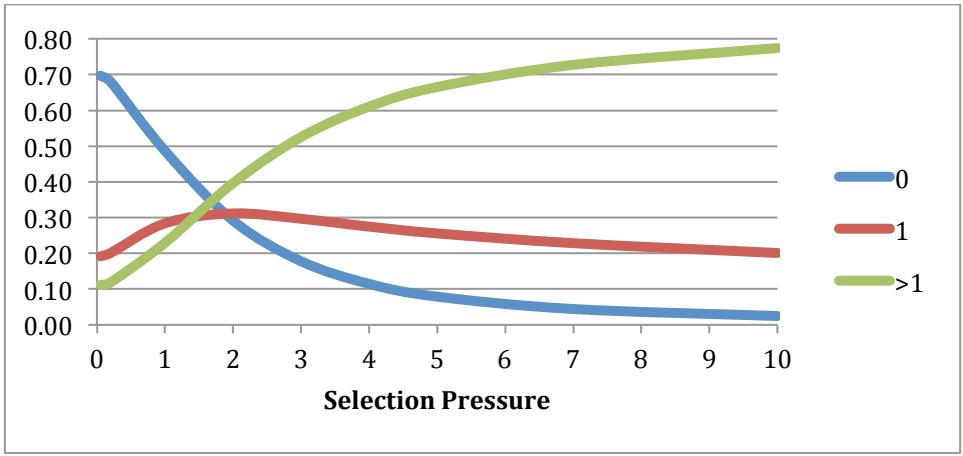

Figure 8. The fraction of non-Kingman collisions (green curve) increases rapidly with selection pressure. Blue, red, green curves are 0, 1, >1 changes in lineage number per generation.

**3.5.2 Kinetics of multiple coalescence events.** As noted earlier, the finite state model describes a time epoch that spans only a few mean coalescent times, perhaps 10 - 50 generations at small population size, and can give a detailed view of the kinetics of coalescence during this time. For example, the change in number of lineages, characterized on an over-all basis in **Fig. 9A**, can be resolved into time intervals (**Fig. 9B**). Multiple lineage coalescences per generations (multicolor curves) are expected in the early phases while only unitary changes (red & blue curves) are expected at later stages of the coalescence process.

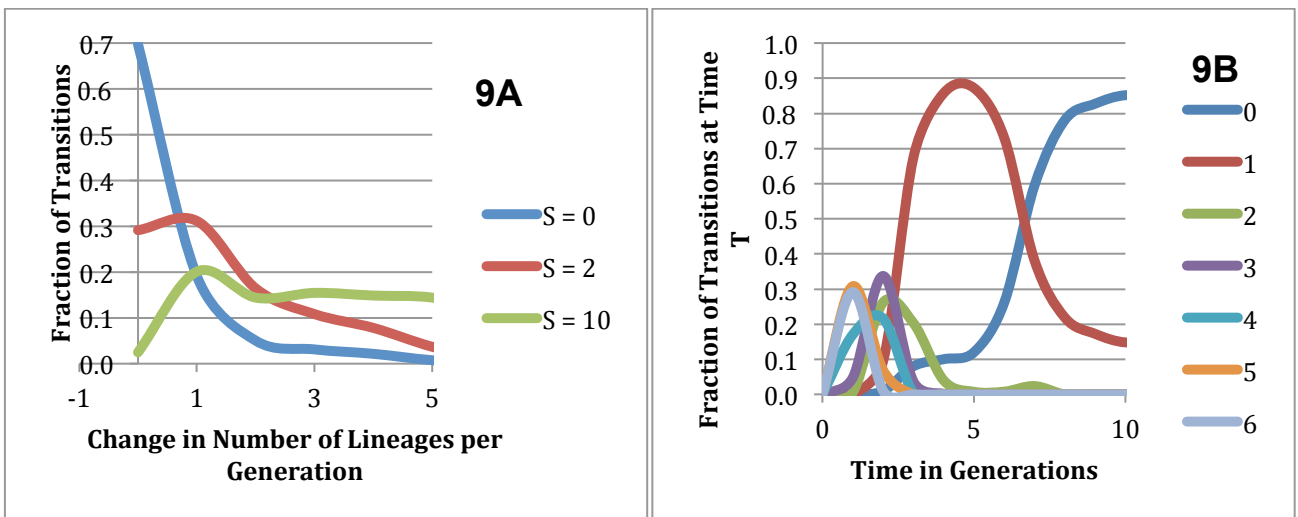

Figure 9. Multiple collisions. Panel 9A: The multiplicity of collisions increases with selection pressure. (N = 10 and selection pressures of 0,2,10). Panel 9B: The multiple mergers occur during early phases at times less than the mean coalescence time. (Number of collisions from 0-6 per generation is color coded).

**3.5.3 Relative Length of Branches.** In the Kingman model, the distribution of branch lengths in a coalescent is given by 2/(i*(i-1)) where **i** is the number of entities still to merge. The last coalescence step accounts for approximately half of the tree height (Wakeley, 2009). Selection pressure reduces the time required for the final coalescence disproportionately, lengthening the relative time consumed during the early stages of coalescence (Neher & Hallatschek, 2013). I see the same effect in the finite state calculations, most easily demonstrated by replotting lineage count vs. time. In **Fig. 10** I normalize the horizontal axis of **Fig. 4**, by the mean coalescent time. The Kingman and overlapping neutral selection finite state curve (blue) show that half the lineages have coalesced in ~ 10% of the mean coalescence time, while systems under selection take up to 30% of the scaled time for this initial drop in lineage number. That is, the branches near the tips of the tree are relatively longer under selection.

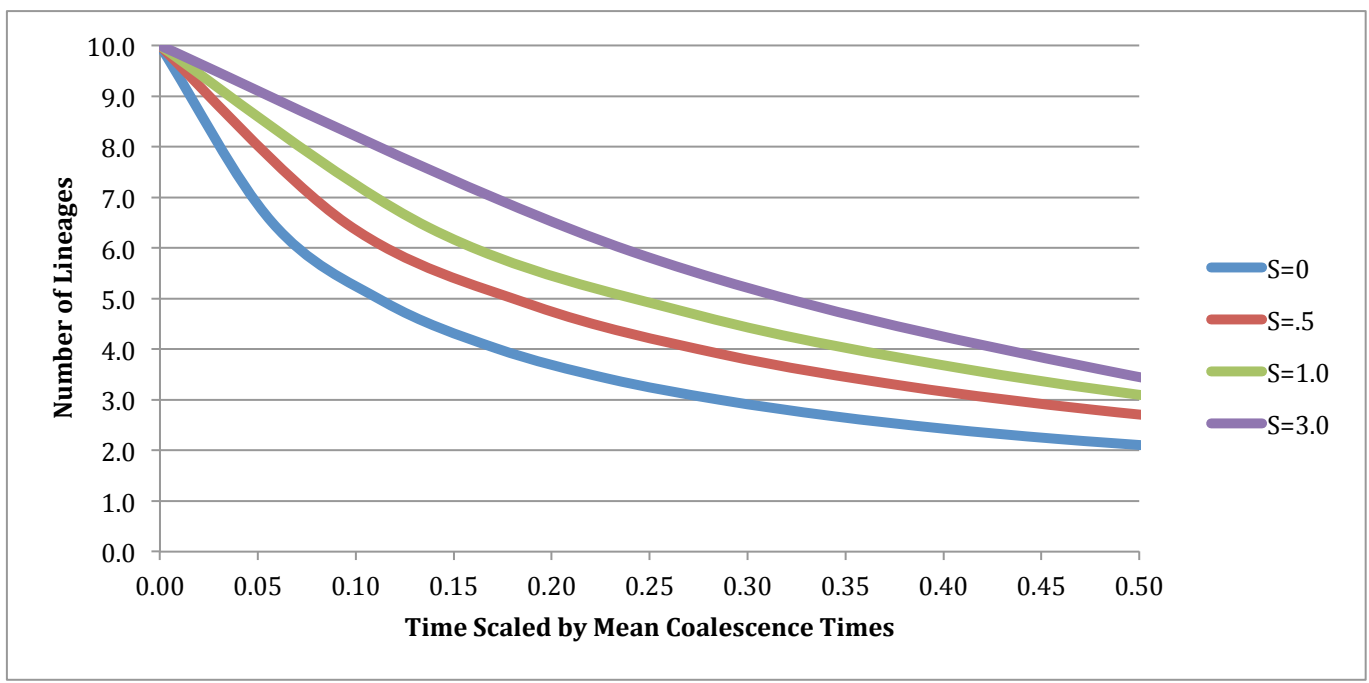

Figure 10. Selection pressure causes slower decline in number of lineages present per scaled time interval. The final coalescence 2 => 1 takes up a smaller portion of the scaled time as selection pressure increases. The finite state data and Kingman coalescent are indistinguishable in the blue curve (**s** = 0).

**3.5.4 Asymmetry of Coalescent Trees.** Aldous has called attention to the asymmetry or "unbalanced" character of coalescence trees observed in both simulations and experimental data (Aldous, 2001). Asymmetry at a binary branch point can be defined as the ratio of the smaller number of progeny on one side of a branch point to the number of progeny on the other side of the branch point. For branch points with more than two descendant clades, the ratio is taken between the smallest clade and the largest clade. The choice of whether to present the most populous branches to the left or to the right about branch points is, of course, arbitrary (Aldous, 2001). A variety of causes of asymmetry have been suggested including "founder" effects, geographic segregation, population bottlenecks, radiation and rapid population expansion (e.g. Heard & Cox, 2007, Pigot et al., 2010). Neher & Hallatschek note significant asymmetry in trees in their model and attribute it to a feature of the Bolthausen-Sznitman coalescent (Fig. 1B, Neher & Hallatschek, 2013). The finite state model lets us examine such effects across the entire set of trees associated with different partitions. I focus on two aspects: the occupancy of asymmetric partitions and the occurrence of partitions that mark symmetry or asymmetry in the trees themselves.

I define "partition asymmetry" by the occupancy of those partitions that contain only 2 lineages, one of which is the selected lineage. These partitions are of the form $a_i b_{N-i}$. Representative occupancies for **N = 10** at the peak of coalescence are shown in **Fig. 11**. At small selection values, the distribution over partitions is nearly symmetrical around the midpoint (dark blue curve centered at .5 in **Fig. 11.** The partition asymmetry increases rapidly as the selection pressure increases (purple & light blue curves, **Fig. 11**).

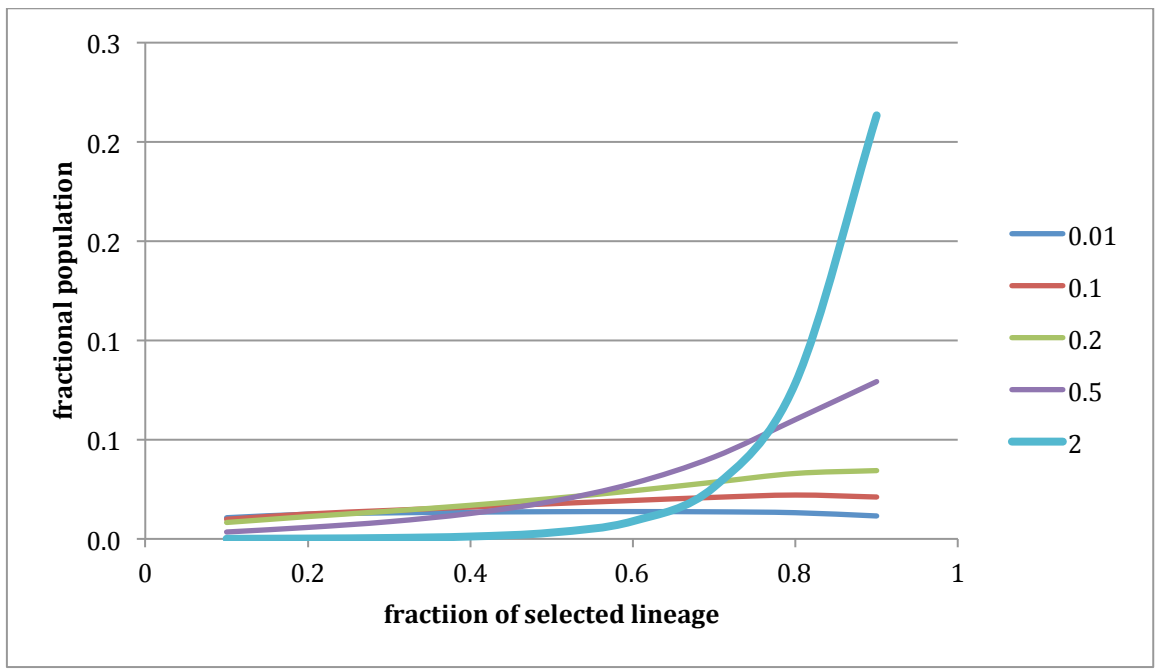

Figure 11. Low selection pressure produces nearly symmetric occupation of 2 lineage partitions with little asymmetry while high selection pressure favors the most asymmetric partitions. 2 lineage partitions of the form $a_i b_{N-I}$, N = 10. Selection pressure from .01 to 2.

I define an asymmetry coefficient by summing the populations of each partition over time and calculating the population-weighted mean value divided by the population size (**Fig. 12A**). This coefficient is 0.5 without selection. It increases with positive selection pressure and decreases when the selection pressure is negative. The asymmetry coefficient depends on population size (**Fig. 12B**). The changes in partition asymmetry appear early in the coalescent process (i.e. near the tips of the branches), when many lineages are coalescing (**Fig. 10**).

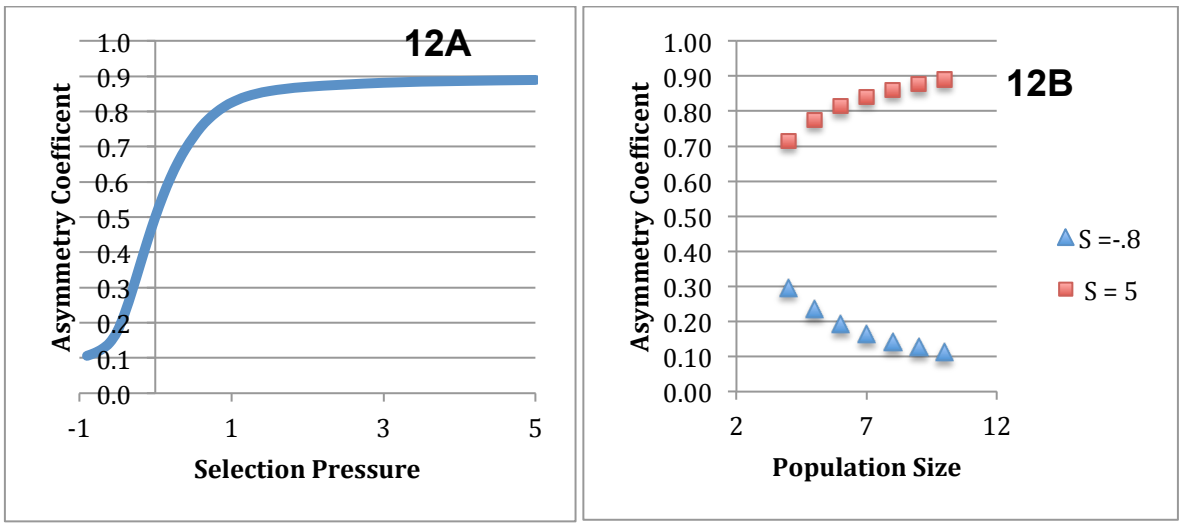

Figure 12. Partition asymmetry coefficient increases rapidly with selection pressure (panel 12A) and population size (panel 12B) at constant selection pressures (**s** = -.8, **s** = 5)

As we have just seen, the most asymmetric partitions are preferentially occupied as selection increases. Now we need to ask if selection drives **individual** trees to assume a more unbalanced shape? Using the simulations we can inspect individual lineage trees and ask two questions. 1) Are the most unbalanced partitions (e.g., for N = 10: {9,1}, {8,1,1}, (7,1,1,1} found to a greater extent in individual trees as selection pressure increases. Second, are balanced partitions such as {5,5}, representing strong bifurcations in the tree, less likely to be seen with selection pressure. The results are shown in **Fig. 13**. Both expectations are met: balanced bifurcations drop and partitions associated with strongly unbalanced trees increase as selection pressure increases. Thus, selection pressure generates increasingly unbalanced individual lineage trees.

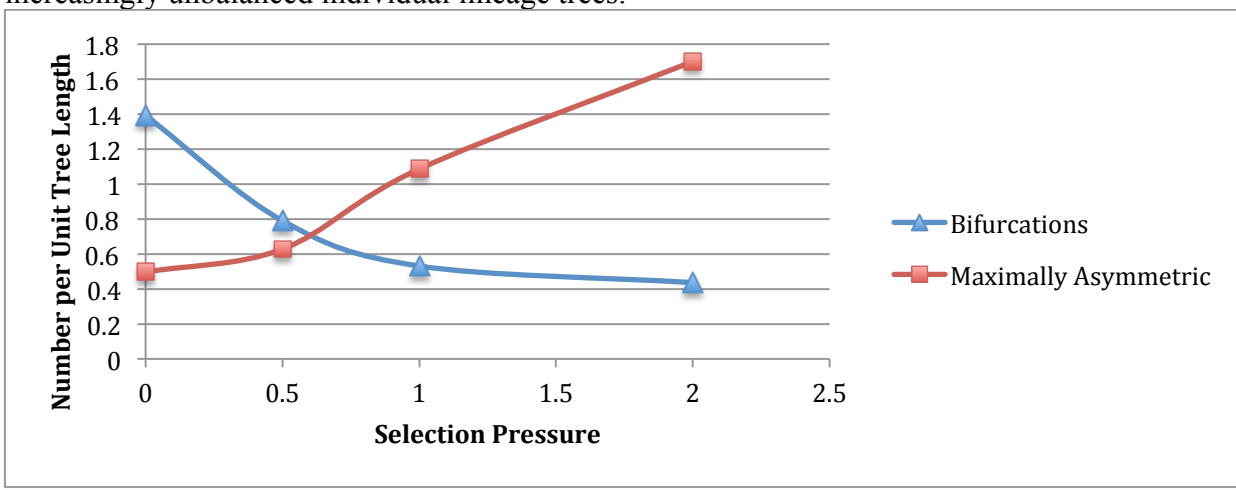

Figure 13. Markers of Symmetric Trees (e.g., partitions with major bifurcations such as {N/2,N/2}) decrease, and of Markers of Maximally Asymmetric Trees (e.g. partitions such as {N-1,1}) increase with selection pressure.

### 3.6 Fixation of Favored Lineages under Selection.

There have been many experimental and theoretical studies of the effect of selection pressure on the fixation of selected alleles (e.g., Haldane, 1927; Fisher, 1937; Desai et al., 2013; Neher & Hallatschek, 2013). The finite state model at small population sizes provides conditions where both selected and non-selected lineages can reach coalescence at each selection pressure (**Fig. 14**). The observed behavior is approximated by the well-known Malécot formula for the Wright-Fisher model (e.g. see Hartl, 2007; Nielsen & Slatkin, 2013).

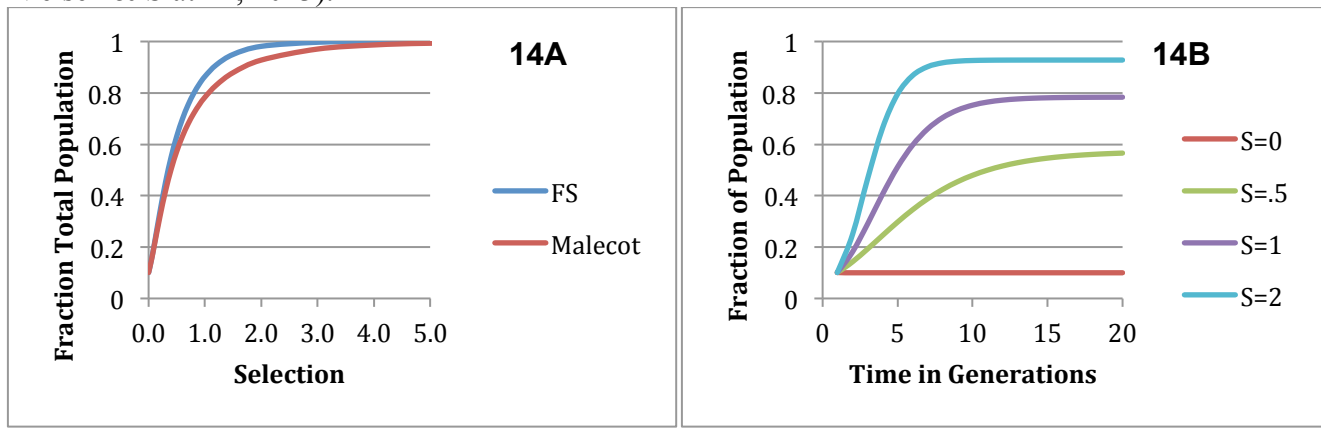

Figure 14. Fraction of total coalescences belonging to selected lineage (A panel, blue curve) is approximately given by the Malécot formula for the Wright-Fisher Haploid Model, $N = 10$. Panel B. Time course for lineage fixation depends strongly on selection.

## 4 Discussion

I have argued that a finite state model provides insight into coalescence phenomena under selection pressure. Selection is treated as a simple shift in the odds of survival for one lineage and my results largely recapitulate those obtained from other theoretical treatments (e.g. Neher & Hallatschek, 2013). Specifically, selection increases the chances of multiple coalescences within a single generation, causes a relative increase in the terminal branch lengths, and increases the number of unbalanced trees and their degree of asymmetry about branch points. The ability to look at individual partitions allowed me to examine non-selected lineages in populations where one lineage had a selective advantage or disadvantage. In the finite state model, the coalescent behaviors of selected and non-selected lineages are almost completely uncoupled. The model can also be extended to "negative selection" regimes. Negative selection strongly decreases the coalescence time of the negatively selected lineages through depletion of the negatively selected lineage population at longer times. It has relatively little impact on the coalescence of the unselected lineages.

Most of these observations are consistent with the Bolthausen-Sznitman coalescent, even though the lineages in the finite state model are not set up as exchangeable. Two criteria used by Neher & Hallatschek, 2013 are not observable in our system: the non-monotonic site frequency spectrum is not available without a mutation mechanism and the pair-wise coalescence distribution is not available with only lineage information.

There has been a long-term interest in extending the Kingman coalescent to include selection. While progress has been made (Wakeley, 2010, 2013), the finite state model highlights the difficulties faced by such efforts, especially the assumptions of neutrality and exclusion of multiple coalescences per generation that are at the heart of the Kingman approach are directly violated by selection. Further, the quantitative issues cannot be removed through use of an $N_{effective}$ term, as also noted by Neher & Hallatschek, 2013. In this regard, the suggestion by Wakeley that the Kingman approach can extend successfully in the face of considerable structure (Wakeley, 2013) has to be approached with caution when selection is involved because the time scale differential can break down with the very large decreases in coalescence times that are possible.

The precise dependence of coalescence on selection suggested by the finite state model resists a simple closed form solution. The limitation of the finite state model to small population sizes will likely remain an issue because of the combinatorial growth in sub-states as the population size increases. However, it might be directly applicable to ecological niches (e.g., Winemiller et al., 2015, Chase & Myers, 2011, White & Adami, 2004).

An open question is whether the relationship of selection pressure and the asymmetry of phylogenetic trees could be inverted to provide some estimates of selection on paleontological time scales from published trees or fossil diversity data (e.g. paleobiodb.org). Currently, there are too many contributing factors to make a straightforward calculation.

## 5 Experiments

Experiments to study coalescent times directly would most naturally make use of microorganisms: bacteria, yeast, and, possibly some viral systems. High-throughput genomics and fitness approaches would be essential. Desai has recently reviewed many of the technical issues (Desai, 2013). If we concentrate on experimental conditions close to those considered here, three challenging tasks must be undertaken. The first is to generate clusters of small populations (10-100 individuals), each capable of being maintained at constant population for multiple generations. While today's microfluidic technology can generate multiple isolated clusters of such size (Abate et al., 2013; Lim & Abate, 2013), achieving constant population is difficult (Moffitt et al., 2012), perhaps requiring intervention on a chamber-by-chamber basis. Even if this problem is solved, one must then be able to analyze each cluster to determine the distribution of lineages - presumably using markers (Eastburn et al, 2013; Lan et al, 2016); Third, one would need a method of introducing and measuring differential fitness (Reynolds et al, 2011; Wiser & Lenski, 2015), ideally in some continuous manner, perhaps through a physical parameter.

## 6 Summary and Conclusions

In the absence of selection, the finite-state model parallels Kingman coalescence theory with only minor model-dependent quantitative differences. The inclusion of selection in a simple way allows inspection of the full range of selection pressures from negative values to high positive values in a continuous manner. The main features of increased selection observed with other approaches are recapitulated here: multiple coalescences per generation, shorter coalescence times, narrower distributions, longer terminal branch lengths, and shifts to log(**N**) dependence of coalescence times on population size at high selection pressure. Selection has a strong influence on tree asymmetry, with features such as long singleton branches arising in a natural way.

The model leads to exact algebraic solutions as a function of selection pressure and population size for many of these features but the large number of intermediate states makes it difficult to reduce these equations to simple closed form treatments, or to extend the model, in its present formulation, to large populations. However, it is feasible to expand the model to include such topics as migration, mutation, distributed selection pressure within a lineage, and multiple lineages with selection. It is also possible to examine other reproduction and population control protocols including allowing the population size to vary within the same framework.

The finite state model exposes one of the inherent difficulties in adding selection to the Kingman coalescent – the assumption of only zero or one lineage reductions per generation is inconsistent with the direct effects of selection. As pointed out by Brunet et al (2007) and also by Neher & Hallatschek (2013), the Bolthausen-Sznitman coalescence describes many aspects of the system we have studied, even though we clearly do not have exchangeable lineages. Nor do we restrict the model to single occurrences of multiple collisions per generation. We also treat small population sizes rather than the large populations in other work. The critical feature of allowing multiple coalescences per generation permits the Bolthausen-Sznitman model to provide at least a semi-quantitative description of a wide range of models of adaptation and selection.

## Acknowledgements

I am most grateful to Igor Rouzine, Ryan Hernandez, and Hao Li of UCSF for very helpful comments. I thank Oskar Hallatschek of UC Berkeley for a discussion of his work. Support was obtained from the Department of Pharmaceutical Chemistry, UCSF, Unrestricted Fund.